\documentclass{article}
\usepackage{spconf,amsmath,graphicx}
\usepackage{amssymb,amsfonts}
\usepackage{breqn}
\usepackage{mathtools}
\usepackage{tikz}
\usetikzlibrary{shapes,arrows}
\usepackage{subcaption}
\usepackage{cite}
\usepackage{xcolor,soul}
\usepackage{hyperref}
\input{mysymbol.sty}
\tikzstyle{phantom vertex} = [ ellipse, 
                               anchor = center, 
                               minimum height = 1*\unit, 
                               minimum width  = 1*\unit,
                               inner sep=0pt,
                               anchor=center]
\tikzstyle{red vertex}   = [black, fill = red!20,   phantom vertex, draw]
\tikzstyle{black vertex} = [black, fill = black!20, phantom vertex, draw]
\tikzstyle{blue vertex}  = [black, fill = blue!20,  phantom vertex, draw]
\tikzstyle{green vertex} = [black, fill = green!20,  phantom vertex, draw]
\tikzstyle{yellow vertex} = [black, fill = yellow!20,  phantom vertex, draw]
\tikzstyle{cyan vertex} = [black, fill = cyan!20,  phantom vertex, draw]
\tikzstyle{vertex}       = [draw, phantom vertex]

\tikzstyle{point} = [ellipse, inner sep=0pt, draw, fill=white, anchor = center,
                     minimum height = 0.05*\unit, minimum width  = 0.05*\unit]

\newtheorem{problem}{\hspace{0pt}\bf Problem}

\title{Robust MIMO Detection using Hypernetworks with Learned Regularizers}
\name{Nicolas Zilberstein$^{\star}$, Chris Dick$^{\dagger}$, Rahman Doost-Mohammady$^{\star}$, Ashutosh Sabharwal$^{\star}$, Santiago Segarra$^{\star}$\thanks{This work was partially supported by Nvidia. Email: \{nzilberstein, doost, ashu, segarra\}@rice.edu, cdick@nvidia.com.}}
\address{$^{\star}$Rice University, USA \hspace{4cm}
         $^{\dagger}$Nvidia, USA}
         
\begin{document}
\ninept
\maketitle
\begin{abstract}
Optimal symbol detection in multiple-input multiple-output (MIMO) systems is known to be an NP-hard problem. 
Recently, there has been a growing interest to get reasonably close to the optimal solution using neural networks while keeping the computational complexity in check. 
However, existing work based on deep learning shows that it is difficult to design a generic network that works well for a variety of channels. 
In this work, we propose a method that tries to strike a balance between symbol error rate (SER) performance and generality of channels. 
Our method is based on hypernetworks that generate the parameters of a neural network-based detector that works well on a specific channel.
We propose a general framework by regularizing the training of the hypernetwork with some pre-trained instances of the channel-specific method.
Through numerical experiments, we show that our proposed method yields high performance for a set of prespecified channel realizations while generalizing well to all channels drawn from a specific distribution.
\end{abstract}

\section{Introduction}
\label{S:introduction}

Multiple-input multiple-output (MIMO) systems are an essential part of modern communications~\cite{mimoreview1}, \cite{mimoreview2}.
Moreover, they are expected to play a fundamental role in moving from the fifth to the sixth generation of cellular communications by achieving high data rates and spectral efficiency~\cite{6g}.
In MIMO systems, base stations are equipped with multiple antennas, enabling them to handle several users simultaneously. %and to use higher modulation rates.
However, these systems entail many challenges such as performing efficient symbol detection, which is the focus of our paper.

Exact MIMO detection is an NP-hard problem~\cite{Pia2017MixedintegerQP}. 
Given $N_u$ users and a modulation of $K$ symbols, the exact maximum likelihood (ML) estimator has an exponential complexity $\mathcal{O}(K^{N_u})$.
Thus, obtaining this ML estimate is computationally infeasible and becomes intractable even for moderately-sized systems. 
Many approximate solutions for symbol detection have been proposed in the classical literature including zero forcing (ZF) and minimum mean squared error (MMSE)~\cite{Proakis2007}. 
Although both (linear) detectors have low complexity and good performance for small systems, their performance degrades severely for larger systems~\cite{chockalingam_rajan_2014}.
Another classical detector is approximate message passing (AMP), which is asymptotically optimal for large MIMO systems with Gaussian channels, but degrades significantly for other (more pratical) channel distributions~\cite{amp}. 

Recently, machine learning and, in particular, deep learning have been proposed to solve fundamental problems in wireless communications such as power allocation~\cite{eisen2020regnn, UWMMSE, chowdhury2021ml}, link scheduling \cite{linkscheduling,zhao2021link}, and random access control~\cite{kumar2021adaptive}.
For the particular case of MIMO symbol detection, several solutions have been derived~\cite{mmnet,detNet2017,hypermimo,remimo}. 
At a high level, one can categorize the existing methods into two classes:
1) \emph{channel-specific methods} learn to perform symbol detection for a prespecified channel realization, and 
2) \emph{channel-agnostic methods} can perform symbol detection for a wide variety of channels, typically drawn from a distribution of interest.
MMNet~\cite{mmnet} and the so-called fixed channel version of DetNet~\cite{detNet2017} are examples of channel-specific methods whereas HyperMIMO~\cite{hypermimo}, RE-MIMO \cite{remimo}, and the varying channel version of DetNet~\cite{detNet2017} are examples of channel-agnostic methods.
Naturally, the first class attains very high performance for the channels in which they were trained but fail in other channels from the same distribution.
However, as they have to be trained for each channel realization, they are typically unsuitable for real-time applications.
In contrast, the second class generalizes well across a distribution without the need of retraining but cannot match the performance of the first class on the specific channels where they were trained.

Our goal is to combine these two classes of methods to attain a solution that yields very high performance for a set of prespecified channels (and their perturbations) and generalizes well to all channels coming from a distribution of interest without the need to retrain. 
We achieve this by first constructing a channel-agnostic methods based on a channel-specific one using the concept of hypernetworks~\cite{ha2016hypernetworks}, and then regularizing the training of the hypernetwork with several pre-trained instances of the channel-specific method.
Although the framework proposed is generic in terms of which channel-specific detector to choose, we focus on MMNet~\cite{mmnet} and its corresponding hypernetwork extension, the HyperMIMO~\cite{hypermimo}.

\vspace{1mm}
\noindent
{\bf Contribution.}
The contributions of this paper are twofold:\\
1) We propose a learning-based solution for MIMO detection that yields high accuracy for perturbations of a prespecified set of channels while generalizing to a whole distribution. We attain this via a HyperMIMO architecture whose training is regularized by solutions of the MMNet.\\
2) Through numerical experiments, we demonstrate that the proposed solution achieves symbol error rates below those obtained by HyperMIMO and MMNet trained separately while maintaining the (forward-pass) computational complexity of HyperMIMO.

\section{System model and problem formulation}
\label{S:system_and_problem}

We consider a communication channel with $N_u$ single-antenna transmitters or users and a receiving base station with $N_r$ antennas. 
The forward model for this MIMO system is given by
\begin{equation}\label{E:mimo_model}
    \bby = \bbH \bbx + \bbn,
\end{equation}
where $\bbH \in \mathbb{C}^{N_r \times N_u}$ is the channel matrix, $\bbn \sim \mathcal{N}(\bb0, \sigma^2 \bbI_{N_r})$ is a vector of complex circular Gaussian noise, $\bbx \in \mathcal{X}^{N_u}$ is the vector of transmitted symbols, $\mathcal{X}$ is a finite set of constellation points, and $\bby \in \mathbb{C}^{N_r}$ is the received vector. 
In this work, a quadrature amplitude modulation (QAM) is used and each symbol is normalized to unit average power. 
It is assumed that the constellation is the same for all transmitters and each symbol has the same probability of being chosen by the users $N_{u}$. 
Moreover, perfect channel state information (CSI) is assumed, which means that $\bbH$ and $\sigma^2$ are known at the receiver.\footnote{To avoid notation overload, we adopt the convention that whenever we assume $\bbH$ to be known, $\sigma^2$ is also known.} 
Under this setting, the MIMO detection problem can be defined as follows.
\begin{problem}\label{P:main}
Given perfect CSI and an observed $\bby$ following~\eqref{E:mimo_model}, find an estimate of $\bbx$.
\end{problem}
Given the stochastic nature of $\bbn$ in~\eqref{E:mimo_model}, a natural way of solving Problem~\ref{P:main} is to search for the $\bbx$ that maximizes the probability of observing our given $\bby$.
Unfortunately, such an ML detector boils down to solving the optimization problem
\begin{equation}
    \hat{\bbx}_{\mathrm{ML}} = \argmin_{\bbx \in \mathcal{X}^{N_u}}\,\, ||\bby - \bbH\bbx||^2_2,
\end{equation}
which is NP-hard due to the finite constellation constraint $\bbx \in \mathcal{X}^{N_u}$~\cite{Pia2017MixedintegerQP}, rendering $\hat{\bbx}_{\mathrm{ML}}$ intractable in practical applications.
Consequently, several schemes have been proposed in the last decades to provide efficient approximate solutions to Problem~\ref{P:main}, as mentioned in Section~\ref{S:introduction}.

The classical body of work (ZF, MMSE, AMP) focuses on solving a single instance of Problem~\ref{P:main} for arbitrary $\bby$ and $\bbH$, which must then be repeated to recompute the detection in successive communication instances.
Given that in practice we are interested in solving several instances of Problem~\ref{P:main} across time, a learning-based body of work has gained traction in the past years~\cite{mmnet, hypermimo, detNet2017, remimo}. 
In a nutshell, based on many tuples $(\bby, \bbH, \bbx)$, the idea is to learn a map -- a function approximator -- from the space of observations and CSI to the corresponding (approximate) transmitted symbols $\bbx$. 
In this way, when a new observation $\bby$ is received (along with the CSI), $\bbx$ can be efficiently estimated using the learned map without the need for solving an optimization problem. 

Having introduced this framework, we can provide a precise distinction between the families of learning-based methods that we denominated as channel-specific and channel-agnostic.
Channel-specific methods like DetNet~\cite{detNet2017} (for the fixed channel case) and MMNet~\cite{mmnet} learn \emph{a different function for every $\bbH$}, i.e., they learn a function $\Phi_{\bbH}:\mathbb{C}^{N_r} \to \mathcal{X}^{N_u}$ such that $\Phi_{\bbH}(\bby)$ is a good solution to Problem~\ref{P:main} for a specific $\bbH$ of interest.
On the other hand, channel-agnostic methods like HyperMIMO~\cite{hypermimo} and RE-MIMO~\cite{remimo} consider the CSI as input to their learnable functions, i.e., they look for $\Phi: \mathbb{C}^{N_r} \times \mathbb{C}^{N_r \times N_u} \to \mathcal{X}^{N_u}$ such that $\Phi(\bby, \bbH)$ is a good solution to Problem~\ref{P:main}.
Naturally, such a satisfactory $\Phi$ cannot be found for completely arbitrary $\bbH$ and, rather, channel-agnostic methods focus on channels drawn from some distribution of interest.
Moreover, due to the specialized nature of $\Phi_{\bbH}$, channel-specific models tend to perform better for the particular channel $\bbH$ but their performance quickly degrades when a different channel is drawn.

In this setting, we are motivated by the following question:
\emph{Can we develop a generalizable channel-agnostic method that achieves performance comparable with channel-specific methods for a channel $\bbH$ (or set of channels $\ccalH$) of interest?}
In essence, we want to keep the best of both classes of methods by performing close to optimal on prespecified channels while generalizing to a whole distribution.
Our motivating question is relevant in practice when, e.g., the channel fading varies smoothly with time as in Jakes model~\cite{Nasir2019} (see Section~\ref{S:numerical_experiments} for more details).
In such a case, we want our learning scheme to perform especially well around the current channel while generalizing satisfactorily to avoid the need for immediate retraining.

%%%%%%%%%%   F   I   G   U   R    E
\begin{figure}[t]
	\centering
	\includegraphics[width=0.8\linewidth]{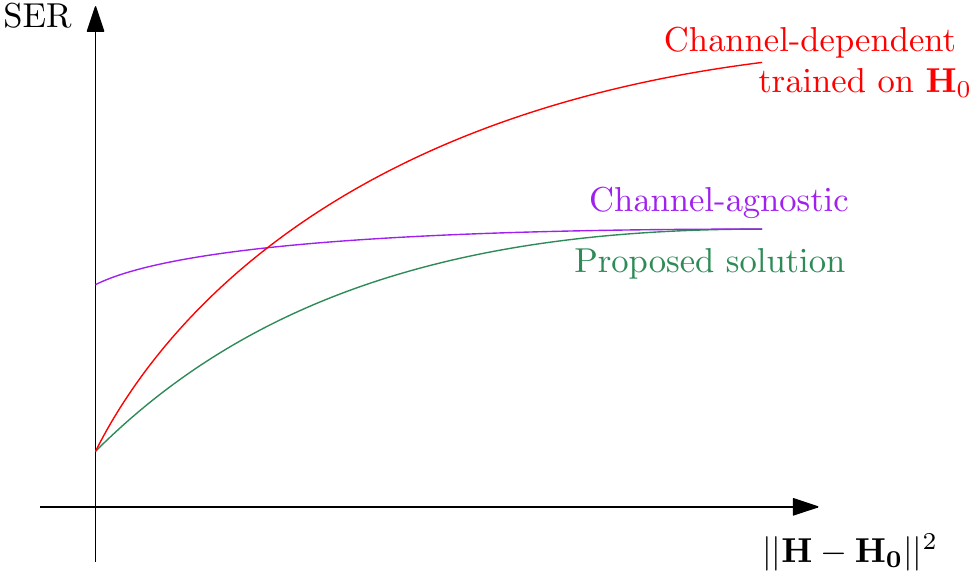}
	\caption{Our proposed solution seeks to combine the best of both classes of methods by specializing to a channel (or set of channels) of interest while generalizing to a whole channel distribution.}
	\label{fig:scheme_error_decrease}
\end{figure}
%%%%%%%%%%%%%%%%%%%%%

At a high level, given some metric in the space of channels, channel-specific solutions yield lower symbol error rate (SER) close to the channel $\bbH_0$ for which they were trained whereas channel-agnostic methods work better when channels further away from $\bbH_0$ are drawn; see Fig.~\ref{fig:scheme_error_decrease}.
Intuitively, we want to derive a method that attains the behavior illustrated in green in Fig.~\ref{fig:scheme_error_decrease}.
Hence, one can think of our sought solution as a \emph{robust} version of a channel-specific method that gracefully degrades into a channel-agnostic method.
Alternatively, one can see the envisioned solution as a channel-agnostic method that has been specially tuned to overperform on a subset of channels of interest.
Either way, we propose to achieve this through the use of hypernetworks whose training is regularized by the solutions of channel-specific methods, as we detail next.

\section{Hypernetworks with Learned Regularizers}
\label{S:hypernetworks_learned_regularizers}

In Section~\ref{Ss:hypernetworks_intro} we introduce the notion of a hypernetwork and its use in machine learning whereas in Section~\ref{Ss:learning_hypernetworks} we detail how we incorporate hypernetworks in our solution to Problem~\ref{P:main}.

\subsection{Hypernetworks in machine learning and MIMO detection}
\label{Ss:hypernetworks_intro}

\begin{figure}[t]
	\centering
	\includegraphics[width=0.75\linewidth]{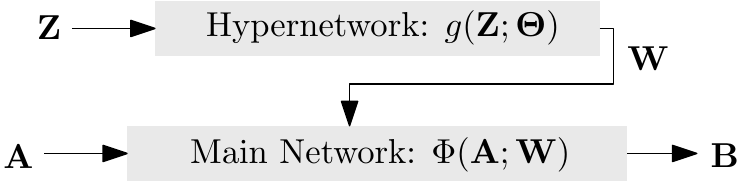}
	\caption{The general scheme of a hypernetwork. The hypernetwork $g(\cdot; \boldsymbol{\Theta})$ takes the input $\bbZ$ and generates the weights $\bbW$, which are fed into the main network. Then, the main network $\Phi(\cdot;\bbW)$ takes as input $\bbA$ and returns the output $\bbB$.}
	\label{fig:hyperNet}
\end{figure}

Hypernetworks are neural networks that have as output the weights of another target (or main) neural network, which performs the learning task~\cite{ha2016hypernetworks}; see Fig.~\ref{fig:hyperNet}.
More precisely, the interpretation of our main network is that of a classical neural network that learns a parametric function $\Phi(\cdot; \bbW):\ccalA \rightarrow \ccalB$ from input data $\bbA$ to some desired target $\bbB$, being $\bbW \in \ccalW$ the learnable parameters of this neural network.
The goal of the hypernetwork, on the other hand, is to learn a parametric function $g(\cdot;\boldsymbol{\Theta}):\ccalZ \rightarrow \ccalW$ from the (possibly different) input $\bbZ$ into the space of parameters $\ccalW$ of the main network.
In this way, we are not fixing the parameters $\bbW$ of our main network but rather making these a function of the input $\bbZ$, effectively improving the generalizability of $\Phi$.
Thus, given inputs $(\bbA, \bbZ)$, the output of the main network is given by $\Phi(\bbA; g(\bbZ;\boldsymbol{\Theta})) \in \ccalB $.
It should be noted that only the parameters $\boldsymbol{\Theta}$ of the hypernetwork need to be learned during training.

%%%%%%%%%%   F   I   G   U   R    E
%%%%%%%%%%%%%%%%%%%%%

The notion of a hypernetwork has been used in several contexts such as object recognition~\cite{Bertinetto2016LearningFO} and generation of 3-D point clouds~\cite{spurek2020hypernetwork}.
For example, hypernetworks have been used for 3D shape reconstruction~\cite{littwin2019deep} and to learn shared representations of images~\cite{sitzmann2020implicit}.
In the specific context of MIMO detection, the use of hypernetworks has been already proposed in \cite{hypermimo} applied to the MMNet.
As the MMNet depends on a particular channel realization, the hypernetwork enables the generalization to a whole distribution of channels.

\subsection{Learning hypernetwork regularizers}
\label{Ss:learning_hypernetworks}

\begin{figure}[t]
	\centering
	\includegraphics[width=0.9\linewidth]{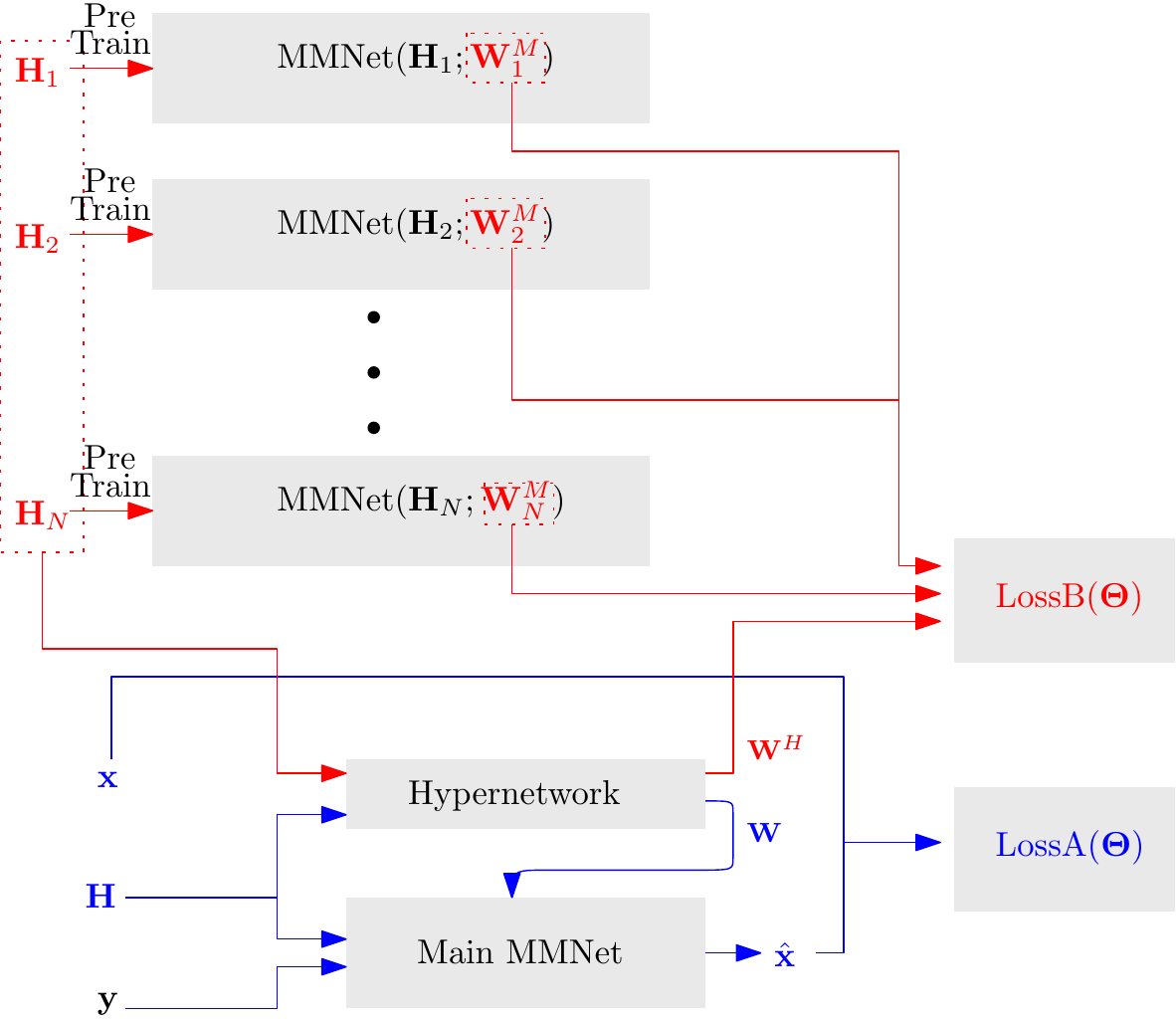}
	\caption{Scheme of our proposed training architecture. 
	In addition to the more classical hypernetwork training (blue arrows), we propose a regularization term that depends on several pretrained main (MMNet) networks, as depicted by the red arrows.}
	\label{fig:scheme_reg}
\end{figure}

Having formally introduced the concept of a hypernetwork, we can now revisit the HyperMIMO~\cite{hypermimo}, which exactly follows the framework in Fig.~\ref{fig:hyperNet}.
In particular, we have that the main network -- which takes the form of an MMNet~\cite{mmnet} -- has as input the observation $\bby$ and the CSI, i.e., $\bbA = \{ \bby, \bbH\}$.
Moreover, the hypernetwork takes the CSI as input $\bbZ = \bbH$ and generates the weights for the multiple layers of the main MMNet.
Then, the MMNet generates the estimate of the transmitted symbols ($\bbB = \hat{\bbx}$).
The weights of the hypernetwork are trained to minimize a loss that compares $\hat{\bbx}$ with the true transmitted symbols $\bbx$.
This flow is depicted by blue arrows in Fig.~\ref{fig:scheme_reg}.

We expand the described training procedure to attain a solution to Problem \ref{P:main} that captures the desirable behavior in Fig.~\ref{fig:scheme_error_decrease}. 
First, we determine the channel or set of channels $\ccalH = \{\bbH_1,\bbH_2, \cdots, \bbH_N\}$ on which we want our solution to achieve especially high detection performance.
This choice will be guided by the nature of the system where we anticipate that our detector will be deployed.
Given many realizations $(\bby, \bbH_i, \bbx)$ for the channels $\bbH_i \in \ccalH$, we train a collection of MMNets $\Phi_{\bbH_i}( \cdot ;\bbW_i^{\mathrm{M}})$, one per channel $\bbH_i$.
Notice that, given the channel-specific nature of $\Phi_{\bbH_i}$, the learned weights $\bbW_i^{\mathrm{M}}$ entail good detection performance for the channel $\bbH_i$.
We use these pretrained weights as regularizers during the training of our hypernetwork; see red arrows in Fig.~\ref{fig:scheme_reg}.
To be more precise, if we denote by $\bbW_i^{\mathrm{H}} = g(\bbH_i; \boldsymbol{\Theta})$ the weights output by the hypernetwork when we input channel $\bbH_i$, we define our regularized loss as
\begin{align}\label{eq:loss}
    \ccalL(\boldsymbol{\Theta}) &= \underbrace{\mathbb{E}_{\bbH,\bbx,\bbn}[||\bbx-\hat{\bbx}||_2^2]}_{\text{LossA}(\boldsymbol{\Theta})} + \underbrace{\beta\sum_{\bbH_i \in \mathcal{H}} ||\bbW_{i}^{\mathrm{M}} - \bbW_{i}^{\mathrm{H}}||_1}_{\text{LossB}(\boldsymbol{\Theta})}.
\end{align}

%%%%%%%%%%   F   I   G   U   R    E
%%%%%%%%%%%%%%%%%%%%%

The first term in~\eqref{eq:loss} computes a classical mean square error between the true symbols and the estimated symbols, where the expected value is taken over the channel, input, and noise distributions of interest.
The channel distribution used here is the one for which we want our channel-agnostic method to generalize.
The second term penalizes the distance between the parameters in the pretrained MMNets and those generated by the hypernetwork when it is fed with channels from $\ccalH$.
We measure this discrepancy using an $\ell_1$ norm to promote a sparse difference between $\bbW_{i}^{\mathrm{M}}$ and $\bbW_{i}^{\mathrm{H}}$.
This means that the weights $\bbW_{i}^{\mathrm{H}}$ generated by the hypernetwork tend to coincide with $\bbW_{i}^{\mathrm{M}}$ for a subset of the entries.
The relative weight $\beta$ captures the importance of performing well within the set $\ccalH$.
Indeed, when $\beta=0$ our proposed method boils down to HyperMIMO and completely ignores the prespecified channels $\ccalH$.
On the other hand, for $\beta \to \infty$ (and assuming that the hypernetwork is sufficiently expressive) our method should mimic the behavior of MMNet on $\ccalH$ but quickly degrade for generic channels.
By selecting an intermediate value of $\beta$, we can realize the desired behavior in Fig.~\ref{fig:scheme_error_decrease}, as we demonstrate in Section~\ref{S:numerical_experiments}.
The model is trained by minimizing the loss in~\eqref{eq:loss} with respect to the hypernetworks parameters $\boldsymbol{\Theta}$ through stochastic gradient descent.

Before presenting the numerical experiments, two remarks are in order.
First, although for concreteness we present our scheme in Fig.~\ref{fig:scheme_reg} for the case where the main network is an MMNet, the same framework can be applied for \emph{any generic channel-specific method} taking the role of our main network.
Second, although our proposed scheme incurs an additional training load in comparison with a vanilla hypernetwork, once trained their computational complexities for detection are exactly the same.
We will refer to our proposed methodology particularized for the MMNet as HyperMIMO with learned regularizers, or HyperMIMO-LR for short. 

%%%%%%%%%%   F   I   G   U   R    E
\begin{figure*}[t]
	\begin{subfigure}{.33\textwidth}
		\centering
		\includegraphics[width=0.95\textwidth]{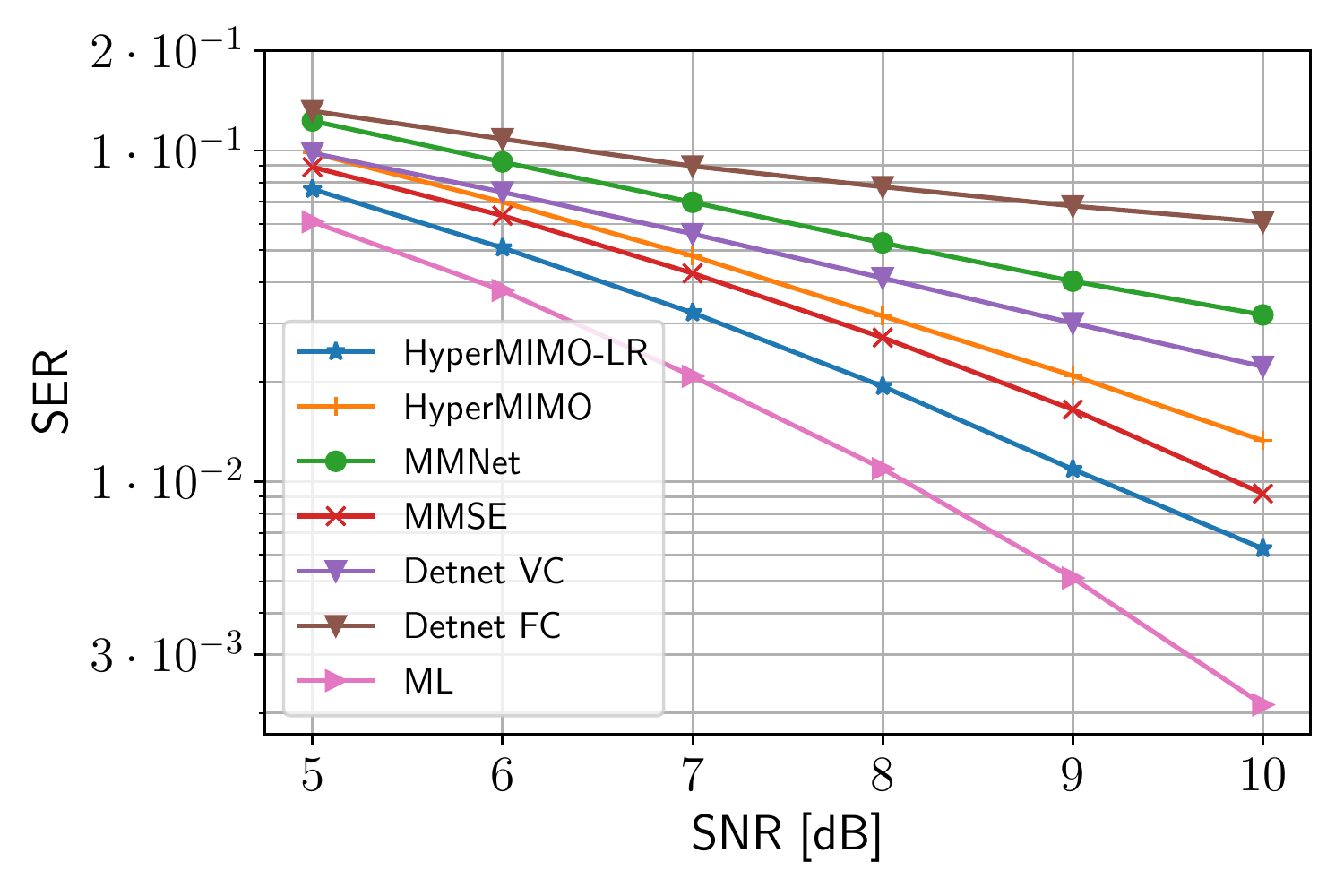}
		\vspace{-0.08in}
		\caption{}
		\label{fig:ser}
	\end{subfigure}%
	\begin{subfigure}{.33\textwidth}
		\centering
		\includegraphics[width=0.95\textwidth]{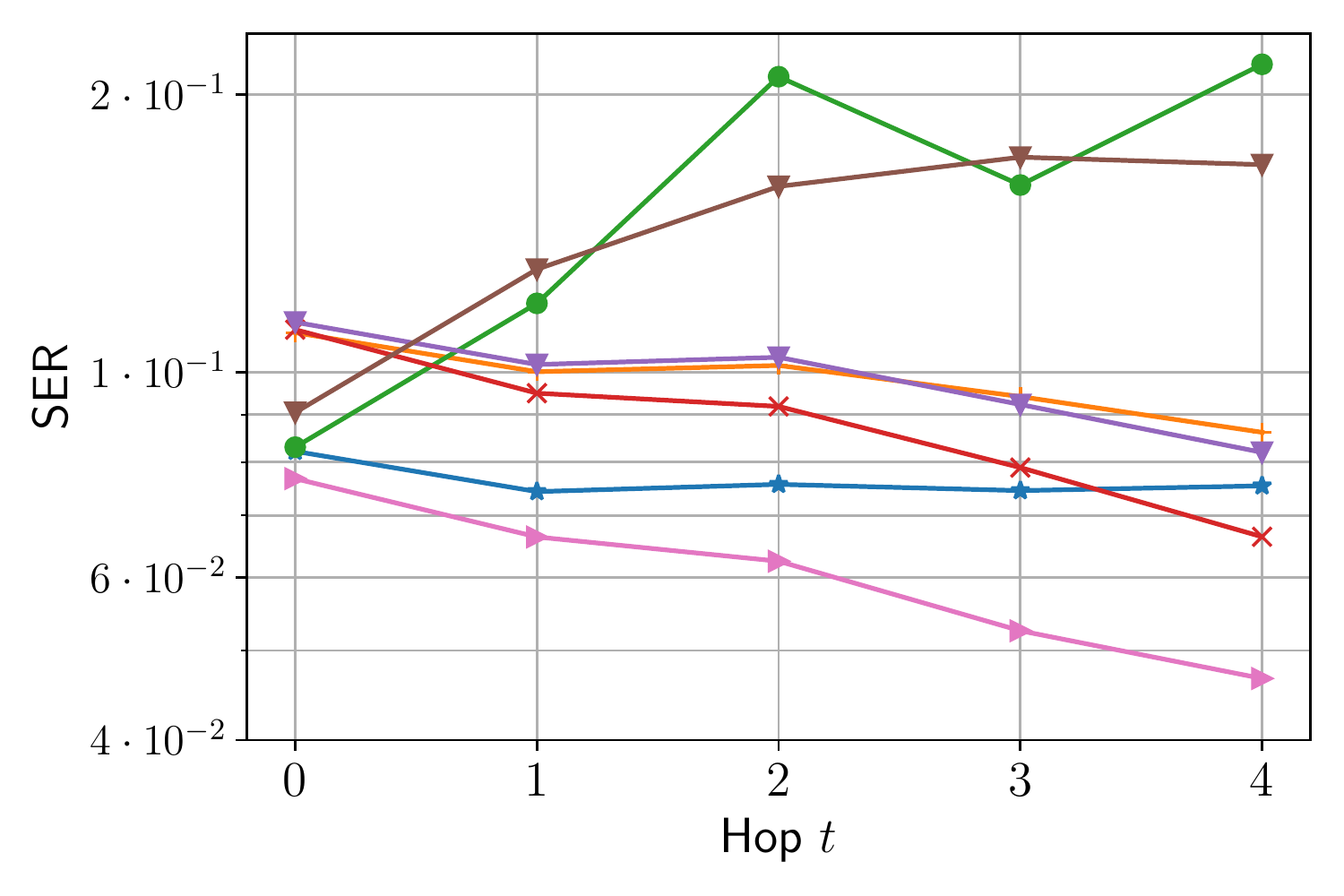}
		\vspace{-0.08in}
		\caption{}
		\label{fig:hops_snr5}
	\end{subfigure}
	\begin{subfigure}{.33\textwidth}
		\centering
		\includegraphics[width=0.95\textwidth]{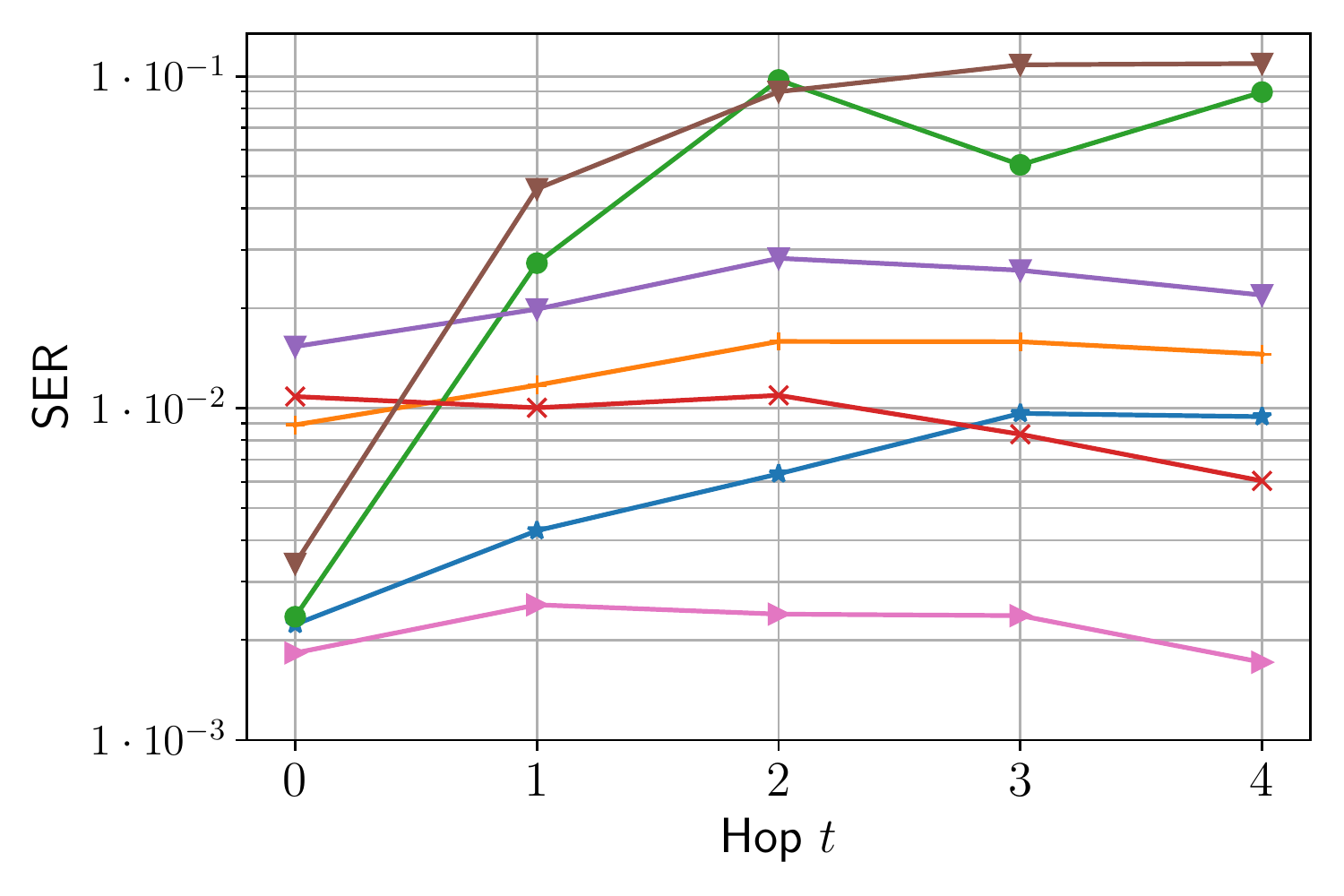}
		\vspace{-0.08in}
		\caption{}
		\label{fig:hops_snr10}
	\end{subfigure}%
	\vspace{-0.05in}
	\caption{(a)~SER as a function of SNR for different detection methods evaluated in a set of channel sequences generated via~\eqref{eq:channel_seq}. (b and c)~SER as a function of time for SNR = $5\text{dB}$ and SNR = $10\text{dB}$, respectively (the same legend as Fig.~\ref{fig:ser} holds).}
	\vspace{-0.1in}
	\label{fig_results}
\end{figure*}
%%%%%%%%%%%%%%%%%%%%%
% \clearpage
\section{Numerical Experiments}
\label{S:numerical_experiments}

In this section we present the results of our proposed method\footnote{Code to replicate the numerical experiments can be found
at \url{https://github.com/nzilberstein/HyperMIMO_LR.git}.}. 
We start by presenting the channel model, simulation setup, and neural network training process.
Then, we present the experimental results and derive insights into the performance of the HyperMIMO-LR.

\subsection{Channel model}

The channel model is generated following the Jakes model \cite{Nasir2019}. 
We express the small-scale Rayleigh fading component as a first-order complex Gauss-Markov process
\begin{equation}
    \bbH_{t} = \rho \bbH_{t-1} + \sqrt{1-\rho^2}\bbe_{t},
    \label{eq:channel_seq}
\end{equation}
where $\{\bbe_{0}, \bbe_{1},\cdots\} \sim \ccalC\ccalN(0,\bbI_{N_r})$ are independent and identically distributed circularly symmetric complex Gaussian random variables.
The initial matrix $\bbH_{0}$ is generated following the Kronecker correlated channel model
\begin{equation}
    \bbH_0 = \bbR_r^{1/2}\bbH_e \bbR_u^{1/2},
    \label{E:kron}
\end{equation}
where $\bbH_e \sim \ccalC\ccalN(0,\bbI_{N_r})$ and $\bbR_r$ and $\bbR_u$ are the spatial correlation matrices at the receiver and transmitter, respectively, generated according to the exponential correlation matrix model with a correlation coefficient $\rho_k$ \cite{Loyka2001}.
In our model, the signal-to-noise ratio (SNR) is given by 

\begin{equation}
    \text{SNR} = \frac{\mathbb{E}[||\bbH\bbx||^2]}{\mathbb{E}[||\bbn||^2]} = \frac{N_u}{\sigma^2N_r}.
\end{equation}
For the experiments, SNRs between $5\text{dB}$ and $10\text{dB}$ are considered.
\subsection{Implementation}

Our simulation environment includes a base station with $N_r=4$ receiver antennas and $N_u=2$ transmitting single-antenna users. We consider 4-QAM modulation. 
The architecture of the hypernetwork is composed of three dense layers: the first layer has the same number of units as the input, the second one has 100 units and the third one has the number of units matching the number of parameter that MMNet requires.
For the MMNet, we use 6 layers.
The activation function for all layers in the hypernetwork is an ELU function; the reason why using an ELU and not a ReLU resides on the nature of the $\boldsymbol{\Theta}$ parameters, which can take negative values.

\noindent \textbf{Training.} We use a batch size of 100 channel matrices generated from~\eqref{E:kron}.
The training is performed using ADAM optimizer \cite{kingma2017adam} with a reduce plateau scheduler: we compute the loss every 500 iterations and when the loss stopped improving, the learning rate is reduced by a factor of $0.9$. 
We train for $50,\!000$ iterations\footnote{For HyperMIMO we followed the same scheme as in \cite{hypermimo}, changing only the lower limit to $10^{-6}$.}.
The value of $\beta$ in~\eqref{eq:loss} was set to $1$.
For the regularizer, we generate $140$ different sequences of length $t=4$ following~\eqref{eq:channel_seq} with $\rho=0.98$ and starting from the same initial matrix $\bbH_0$ from~\eqref{E:kron} with $\rho_k = 0.6$.
In total, we use $N = 561$ pre-trained MMNets.

\subsection{Simulation results}

For testing the performance of the detectors, we generate a test set of $100$ sequences of the same length $t=4$ from the same model in~\eqref{eq:channel_seq}, also starting from $\bbH_{0}$.

We compare the SER achieved by HyperMIMO-LR with respect to the following methods: HyperMIMO, MMNet, DetNet with fixed and varying channel, MMSE and ML (using the Gurobi solver \cite{gurobi}).
The comparisons are shown in Fig. \ref{fig:ser}.
The figure reveals that the performance of HyperMIMO-LR is closest to the optimal ML, and outperforms all the other methods, in particular both HyperMIMO and MMNet. 
It is particularly interesting to observe that while HyperMIMO-LR consistently outperforms the classical MMSE detector, HyperMIMO has a worse performance than MMSE. This is because the performance of the HyperMIMO decreases significantly when it is tested in perturbed versions of a channel from the distribution, while HyperMIMO-LR performs robustly in those unseen channels.

The performance of the detector as a function of the hops $t$ is represented in Figs. \ref{fig:hops_snr5} and  \ref{fig:hops_snr10}, for an SNR of $5\text{dB}$ and $10\text{dB}$, respectively.
We use the same test set as in the previous experiment.
In both cases, we observe equal SERs at $t=0$ (initial matrix $\bbH_0$) for both MMNet and HyperMIMO-LR.
This is expected because the parameters of both architectures are similar due to the regularizer, and hence the performance of both has to be the same at the initial hop.
We also see that the performance of DetNet-FC at the initial hop is close to HyperMIMO-LR and MMNet, but the performance for both MMNet and DetNet-FC quickly degrades as we increase $t$.
Moreover, HyperMIMO follows a similar trend as HyperMIMO-LR, meaning that its performance does not drop severely with $t$ but nonetheless it is inferior to HyperMIMO-LR.
Overall, this behaviour is what we expected from our motivation defined in Fig. \ref{fig:scheme_error_decrease}.
Lastly, we see that MMSE performs relatively better for later hops $t$.
This can be explained by looking at the Jakes model in~\eqref{eq:channel_seq}: as we get farther from $\bbH_{0}$, the Gaussian component tends to dominate. 
In such a Gaussian regime, MMSE achieves a very good performance.

\section{Conclusions}

We proposed a general deep learning based solution for MIMO detection that achieves a high performance for perturbations of a prespecified set of channels while generalizing to the whole distribution.
This was done by regularizing the training of the hypernetwork to a deep learning-based detector with solutions for a set of specific channels using that detector.
We evaluated this general architecture with an implementation that uses HyperMIMO, a hypernetwork-based solution that incorporates MMNet as its deep learning-based MIMO detector. 
We demonstrated that our implementation, named HyperMIMO-LR, generalizes well to the whole distribution of channels and outperforms HyperMIMO.
Future work include extending to higher-order systems as well as higher-order modulation.
\newpage

\bibliographystyle{IEEEbib}
\bibliography{main}

\end{document}